\documentclass{elsart4}
\usepackage{graphicx}
\usepackage{dcolumn}
\usepackage{bm}
\usepackage{amsmath}
\usepackage{amsfonts}
\usepackage{amssymb}

\begin{document}

\begin{frontmatter}

\title{Carbon nanotube with pressure inducing pseudogaps: Kondo effect study}

\author[IFRJ]{T. Lobo},
\ead{thiago.fonseca@ifrj.edu.br}
\author[UFAM]{Minos A. Neto},
\author[IFAM]{Marcio G. da Silva},
\author[UFAM]{Octavio D. R. Salmon},

\address[IFRJ]
{Instituto Federal de Educa\c{c}\~ao, Ci\^{e}ncia e Tecnologia do Rio de Janeiro\\
Rua Dr. Jos\'e Augusto Pereira dos Santos, s/n (C.I.E.P. 436 Neusa Brizola), 
24425-004 S\~ ao Gon\c{c}alo, Rio de Janeiro, Brasil.} 

\address[UFAM]
{Departamento de F\'{\i}sica, Universidade Federal do Amazonas, 3000, Japiim,
69077-000, Manaus-AM, Brazil}

\address[IFAM]
{Instituto Federal do Amazonas, Av. 7 de Setembro, 1975, Centro, Manaus-AM, Brazil}

\begin{abstract}
In this work we are interested to studying the Kondo effect present in a system with a $T$-shape ligation between a 
single-wall carbon nanotube (SWNT) and a magnetic impurity. The system has been studied under hydrostatic pressure 
and it was observed the opening of the gap in the density of states of the zigzag metallic tube. The pressure can be modeled 
by the Pierls instability and in this work we consider the out-of-plane distortion. A tight-binding approximation is used 
to calculate the SWNT Green's functions with hydrostatic pressure applied. We studied the disappearance of the Kondo peak as 
the gap opens. Moreover, we observed the strong influence of the pressure in the conductance curve that can be explained by 
the variation of Kondo peak height. The Kondo effect was reproduced with the atomic approach with $U\rightarrow\infty$ 
developed previously. Results of the electronic density of states and curves of the conductance are presented.

\end{abstract}


\end{frontmatter}

\section{Introduction}
\label{Sec1}

Since $90's$ decade, when they were discovered, the carbon nanotubes have been receiving attention by good part of the 
scientific community. With unique characteristics, they present a great potential for application, since such areas as biomedicine 
\cite{med1,med2,med3,med4,med5} to computer science \cite{comp1,comp2,comp3}. The experimental evidence of the carbon nanotubes 
came in $1991$ with imaged multi-wall carbon nanotubes (MWNTs) using a transmission electron microscope \cite{Iijima91} and in 
$1993$ the first single-wall carbon nanotubes (SWNTs) were observed \cite{IijimaN93,BethuneN93}.

Throughout the literature, some studies that analyze the pressure effect on the carbon nanotube structure have been object of study 
\cite{capaz2004}. In some experimental works the X-ray diffraction can verify structural changes in SWNT when induced by 
hydrostatic pressure ranging from $0$ to $4,5GPa$. \cite{jamal2013}, but depending on the applied pressure, these structures may 
collapse or undergo large structural modifications (order pressure $\sim$ $13-15 GPa$ \cite{anis2012,aguiar2012} to $\sim$ $25GPa$ \cite{you2011,anis2013}).

More recently, experimental results in SWNT, using low-frequency Raman spectroscopy, have reported critical points of phase 
transitions due to induced deformation \cite{shen2017}. These radial collapse transitions were also observed, using theory and 
experiment, in single-, double- and triple-wall carbon nanotubes \cite{alencar2017}. When we confine hydrogen inside a single wall 
carbon nanotube we can check superconducting properties under high-pressures \cite{nano2019}.

The structure of a SWNT can be described as a graphene sheet rolled into a cylindrical shape so that the structure is one-dimensional 
with axial symmetry. We can define the chiral vector $\vec{C_{h}}$ that is related to the circumference of the tube. The chiral 
vector can be written as a function of the real space unit vetors $\vec{a_{1}}$ and $\vec{a_{2}}$ of the hexagonal lattice as $\vec{C_{h}}=m\vec{a_{1}}+n\vec{a_{2}}$. Here, $m$ and $n$ are the integer indexes defining the three types of the SWNT: the 
zigzag ($m=0,n > 0$), the armchair ($m=n > 0$) and the chiral ($0<\left|m\right|< n$). While the armchair tubes are always metallic, 
the zigzag tubes are metallic when $n$ is multiple of $3$ and semiconductor otherwise.

It is known that the resistivity of a metal with magnetic impurities diluted decreases below a characteristic temperature $T_{K}$ and 
this effect is known as the Kondo effect \cite{Kondo64}. Recent experiments have reported the Kondo effect in metallic carbon 
nanotubes \cite{Jesper2000}. In a previous study, we investigated the Kondo effect considering a SWNT under a magnetic flux which allows an  
insulator-metal transition \cite{TLobo_PB}. 

Now, we use a metallic zigzag SWNT $(3,0)$ to study the interplay between the Kondo effect and the open of the gap when a hydrostatic 
pressure is applied. In the Fig. (\ref{fig1}) we can see a pictorial view of the system that consist in a magnetic impurity laterally 
coupled to a zigzag nanotube.

In Section II we present the model and formalism of Green function (GF). The results and discussions are described in Section III. 
Finally, the last section is devoted to ultimate remarks and conclusions.

\section{Formalism}
\label{Sec11}
The impurity was modeled by the atomic approach \cite{Thiago1_2006qd,Nanotech1} and the GF of the tube is calculated analytically, 
adopting a single $\pi-$band tight-binding Hamiltonian \cite{Thiago2_2006nt}. The effect of a pressure applied perpendicular to 
the axial direction is considered by adopting the Peierls distortion \cite{livro}. Essentially, it amounts to the addition of a 
term into the hopping that represent the distortion of atoms from their original positions. 

The energy dispersion relation $\epsilon^{\pm}(\tilde{k}_{x})$ for the zigzag nanotube, submitted to a deformation in the $z$ plane, 
perpendicular to axe of the tube \cite{livro}, may be written as
\begin{equation}
\epsilon^{\pm}(\tilde{k}_{x})=\pm t \sqrt{\Gamma^{2}+1+4\cos\left(\tilde{q}\right)\cos(\tilde{k}_{x})+4\cos^{2}\left(\tilde{q}\right)}. 
\label{nano8}
\end{equation}%
with $\tilde{k}_{x}=k_{x}a\sqrt{3}/2$, $\tilde{q}=q\pi/n$, $\Gamma=\alpha z_{0}/t$ and for $-\pi/\sqrt{3} < k_{x}a < \pi/\sqrt{3}$; 
$q=1,...,2n$. $D$ is the tube diameter, $a$ is the lattice parameter, $\alpha$ is the electron-fonon coupling constant and $z_{0}$ is 
the out-of-plane deformation.

The GF of zigzag nanotube was previously calculated \cite{Thiago2_2006nt} and can be write by
\begin{equation}
g^{0}_{jj'}(z)=\frac{1}{n}\sum_{l}\frac{a\sqrt{3}}{2\pi}\int^{\pi/a\sqrt{3}}_{-\pi/a\sqrt{3}}
dk_{x}\frac{z.e^{i.\stackrel{\rightarrow}{k}.(\stackrel{\rightarrow}{R_{j}}-\stackrel{\rightarrow}
{R_{j'}})}}{z^{2}-\epsilon^{+}(\tilde{k_{x}})}
\label{gtube}
\end{equation}%
where $l=1,n$ and $k_{y}=2\pi l/an$.

The density of state from the conduction band is calculated using
\begin{equation}
\rho_{c}(z)=-\frac{1}{\pi}Im(g^{0}_{jj'})\end{equation}

\begin{figure}[htbp]
\centering
\includegraphics[width=7.5cm,height=6.0cm,angle=0.0]{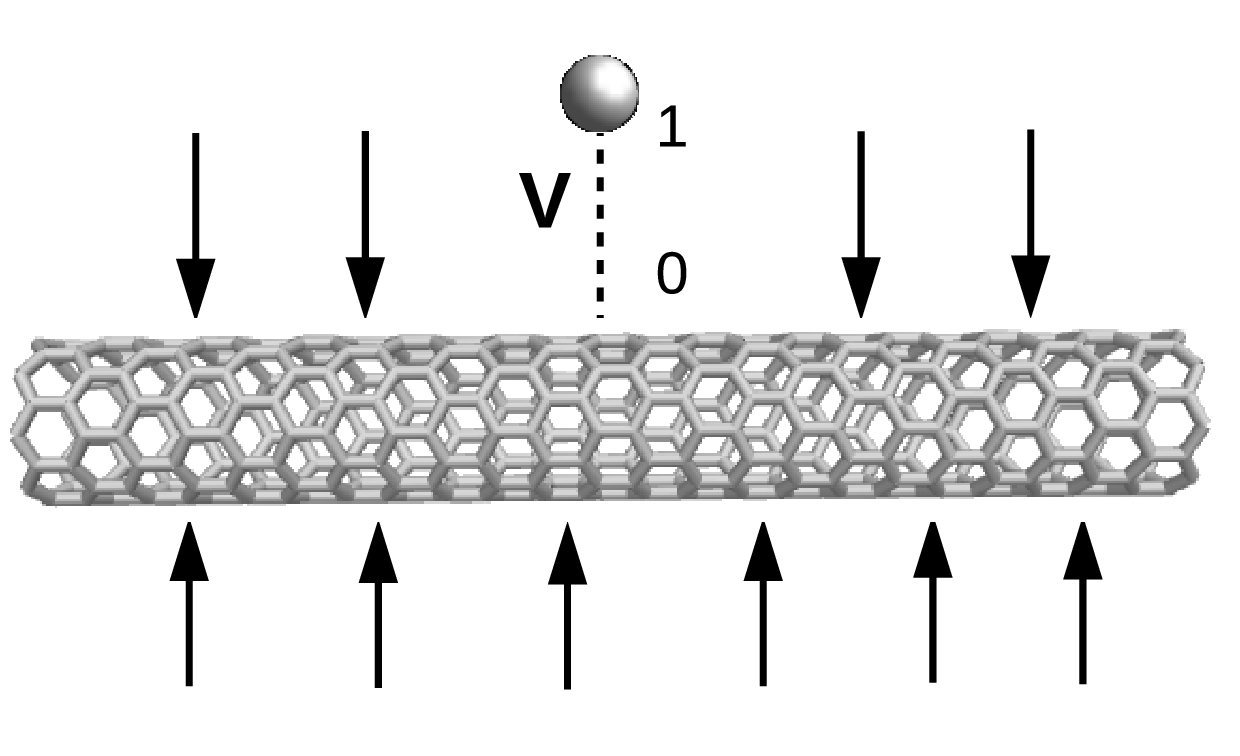}
\caption{Pictorial view of a carbon nanotube with a laterally coupled magnetic impurity in the presence of hydrostatic pressure.} 
\label{fig1}
\end{figure}

We apply the atomic approach\cite{Nanotech1} to study the Kondo effect on a zigzag SWNT with a magnetic impurity side-coupled to its 
surface \cite{QDnosso}, as schematically represented in Fig. (\ref{fig1}).

The Kondo effect was experimentally detected in carbon nanotube devices \cite{Jesper2000} and it investigate the possibility 
of occurrence in the proposed carbon based structure. The system may be modeled by the Anderson impurity model with infinite Coulomb 
repulsion $(U\rightarrow\infty)$. The Hamiltonian is given by

\begin{eqnarray}
\mathcal{H} &=&\sum_{\mathbf{k},\sigma }E_{\mathbf{k},\sigma }c_{\mathbf{k},\sigma}^{\dagger }c_{\mathbf{k},\sigma }+
\sum_{\sigma }\ E_{f,\sigma }X_{f,\sigma\sigma } \nonumber \\
&&+V\sum_{\mathbf{k},\sigma}\left(X_{f,0\sigma}^{\dagger }c_{\mathbf{k},\sigma }+c_{\mathbf{k},\sigma }^{\dagger }X_{f,0\sigma }\right),
\label{Eq.1}
\end{eqnarray}
where the first term represents the conduction electrons ($c$-electrons), the second describes the Anderson impurity characterized 
by a localized $f$ level $E_{f,\sigma}$, (here $f$ indicates localized electrons at the impurity site) and the last one corresponds 
to the interaction between the $c$-electrons and the impurity. For simplicity we consider a constant hybridization $V$. 

Hubbard operators \cite{FFM} are employed to project out the double occupation state $\left|f,2\right\rangle$, from the local states 
on the impurity. The identity decomposition in the reduced space of local states at the impurity is given by $X_{f,00}+X_{f,\sigma\sigma}
+X_{f,\overline{\sigma}\overline{\sigma}}=I$, where $\overline{\sigma}=-\sigma$, and the three $X_{f,aa}$ are the projectors into the 
states $\mid f,a\rangle$. The occupation numbers on the impurity $n_{f,a}=\langle X_{f,aa}\rangle$ should then satisfy the ``completeness'' 
relation

\begin{equation}
n_{f,0}+n_{f,\sigma}+n_{f,\overline{\sigma}}=1.  
\label{Eq.4}
\end{equation}%
To obtain the exact $f$ GF $G_{ff,\sigma}(\mathbf{j}_{i},z)$ in real space for the impurity at site $\mathbf{j}_{i}$, one 
follows a procedure similar to the one used in \cite{FFF} within the chain approximation. The exact GF for the $f$ electron is then 
written as

\begin{equation}
G_{ff,\sigma}(z)=\ \frac{M_{2,\sigma}^{eff}(z)}{1-M_{2,\sigma}^{eff}(z)\mid V\mid^{2}\sum_{\mathbf{k}}G_{c{,}\sigma}^{o}(\mathbf{k}{,}z)\ }.
\label{Eq.6}%
\end{equation}
where $G_{c{,}\sigma}^{o}(\mathbf{k},z)$ is the GF of the conduction electrons of nanotube.

The atomic approach consist to considerate the case of null conduction band, using $U\rightarrow\infty$. In this case the double 
occupation in $f$-electrons is forbidden. We solve the atomic model diagonalizing the matrix ($12\times12$) of the states 
$\left|f,c\right\rangle$, where the possibilities are: $\left|f\right\rangle=0, \uparrow, \downarrow$ and 
$\left|c\right\rangle=0, \uparrow, \downarrow, \uparrow\downarrow$. The GF of the atomic model is obtained by Zubarev equation

\begin{eqnarray}
G^{at}(z) &=& e^{\beta \omega}\sum_{n}\sum_{jj^{\prime}}e^{-\beta E_{n,j}}+e^{-\beta E_{n-1,j^{\prime}}} \nonumber \\
&&\times\frac{\left|\left\langle n-1,j^{\prime}\left|X_{\mu}\right|n,j\right\rangle\right|^{2}}{z-(E_{n,j}-E_{n-1,j^{\prime}})},
\label{Gat1}
\end{eqnarray}%
where $\Omega$ is the thermodynamical potential and the eigenvalues $E_{n,j}$ and eigenvectors $|nj\rangle$ correspond to the 
complete solution of the Hamiltonian. The different transitions occur between states with $n$ and $n + 1$ particles that satisfy 
$\langle n-1, j^{\prime}|X_{\mu}|n,j\rangle\neq 0$. The final result is the following:

\begin{equation}
G^{at}(z)=e^{\beta \omega}\sum^{8}_{i=1}\frac{m_{i}}{z-u_{i}}.
\label{Gat2}
\end{equation}%
The complete calculation of all poles $u_{1}$ and residues $m_{i}$ is detailed in reference \cite{Nanotech1}. On the other hand, 
the exact atomic $f$ GF, calculated exactly within the chain approximation has the same form of Eq. (\ref{Eq.6}) \cite{FFF}: 

\begin{equation}
G_{ff,\sigma}^{at}(z)=\frac{M_{2,\sigma}^{at}(z)}{1-M_{2,\sigma}^{at}(z)\mid V\mid^{2}\sum_{\mathbf{k}}G_{c,\sigma}^{o\bar{e}}(\mathbf{k},z)},\label{Eq.6a}%
\end{equation}
where $G_{c,\sigma}^{o\bar{e}}(\mathbf{k},z)=-1/(z-\varepsilon(\mathbf{k}))$. From this equation we then obtain an explicit expression 
for $M_{2,\sigma}^{at}(z)$ in terms of $G_{ff,\sigma}^{at}(z)$ 

\begin{equation}
M_{2,\sigma}^{at}(z)=\frac{G_{ff,\sigma}^{at}(z)}{1+G_{ff,\sigma}^{at}(z)\mid V\mid^{2}\sum_{\mathbf{k}}G_{c,\sigma}^{o\bar{e}}(\mathbf{k},z)},
\label{Eq.66a}
\end{equation}%
where the exact solution to $G_{ff,\sigma}^{at}$ is given by Eq. (\ref{Gat2}). To decrease the contribution of the $c$-electrons, whose 
effect is overestimated by concentrating them at a single energy level, we replace $V^{2}$ by $\Delta^{2}$, where $\Delta=\pi V^{2}/2D$ 
is of the order of the Kondo peak's width. 

The atomic approach consists in substituting $M_{2,\sigma}^{eff}(z)$ in Eq. (\ref{Eq.6}) by the approximate $M_{2,\sigma }^{at}(z)$ given 
by Eq. (\ref{Eq.66a}). Finally, the local GF of the Anderson impurity is written as

\begin{equation}
G^{imp}_{ff,\sigma}(z)=\ \frac{M_{2,\sigma}^{at}(z)}{1-M_{2,\sigma}^{at}(z)\mid \Delta \mid^{2}\sum_{\mathbf{k}}G_{c,\sigma}^{o}(\mathbf{k},z)},
\label{Eq.61}%
\end{equation}
with $G_{c{,}\sigma}^{o}(\mathbf{k}{,}z)$ given by Eq. (\ref{gtube}).

\section{Results and Discussions}
\label{sec4} 

We are interested in describing the magnetic impurity side-coupled in a metalic zig-zag SWCT. At low-temperatures and 
bias voltage the electronic transport is coherent and the conductance is calculated by Landauer equation \cite{Kang2001}:

\begin{equation}
G=\frac{2e^{2}}{\hbar}\int\left(-\frac{\partial \Theta_{F}}{\partial\omega}\right)  S(\omega)d\omega,
\end{equation}%
$\Theta_{F}$ is the Fermi function and $S(\omega)$ is the transmission probability of an electron with energy $\hbar\omega$ and is 
given by

\begin{equation}
S(\omega)=\Gamma^{2}\mid G_{00}^{\sigma}\mid^{2},
\end{equation}%
where $\Gamma$ corresponds to the coupling strength of the site $0$ to the tube (which is proportional to the kinetic energy 
of the electrons in the site $0$ of the tube). $G_{00}^{\sigma}$ can be calculated by the Dyson's equation, with the hybridization 
operator $\hat{V}=|0\rangle V\langle 1|+|1\rangle V\langle 0|$, see Fig.(\ref{fig1}). 

The dressed GF at the site $0$ can be written in terms of the undressed localized GF $g_{11}$ of the magnetic impurity, and in terms of the 
undressed GF $g_{00}$ of the conduction electrons of the SWNT

\begin{equation}
G_{ii}^{\sigma}=g_{ii}^{\sigma}+g_{ij}^{\sigma}VG_{ji}^{\sigma},
\end{equation}%

\begin{equation}
\nonumber
G_{00}^{\sigma}=g_{00}^{\sigma}+g_{00}^{\sigma}VG_{10}^{\sigma}+g_{01}^{\sigma}VG_{00}^{\sigma},
\end{equation}

\begin{equation}
\nonumber
G_{10}^{\sigma}=g_{10}^{\sigma}+g_{10}^{\sigma}VG_{11}^{\sigma}+g_{11}%
^{\sigma}VG_{00}^{\sigma}.
\end{equation}%
Solving this system of equations, and considering $g_{10}=g_{01}=0$, we can write

\begin{equation}
G_{00}^{\sigma}=\frac{g_{00}^{\sigma}}{(1-g_{00}^{\sigma}V^{2}g_{11}^{\sigma})},
\end{equation}%
where $g_{00}^{\sigma}$ is the undressed GF of the SWNT Eq. (\ref{gtube}) and the $g_{11}^{\sigma}$ is given by the Eq.(\ref{Eq.66a}).

\begin{figure}[htbp]
\centering
\includegraphics[width=7.5cm,height=6.0cm,angle=0.0]{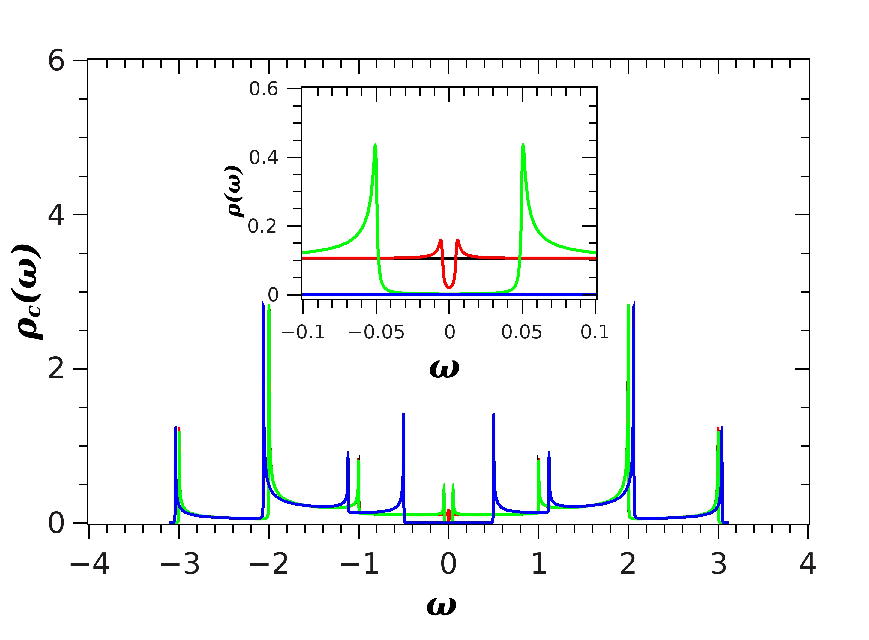}
\caption{Density of states of a (3,0) SWNT for $\Gamma=0.0\Delta$, $\Gamma=0.5\Delta$, $\Gamma=5.0\Delta$ and $\Gamma=50.0\Delta$.} 
\label{fig2}
\end{figure}

In the Fig. (\ref{fig2}0 we show the DOS of the $c$-electrons of the $(3,0)$ zigzag nanotube, where $\rho_{c}=-\frac{1}{\pi}\Im(g_{00}^{\sigma})$. 
Note the DOS structure which features a set of several peaks, known as Van Hove singularities. For the case $\Gamma =0$, we 
observe that the nanotube is metallic, as expected. As we increase the $\Gamma$ value, a pseudogap arises in the chemical 
potential $\mu=0$. For extreme values of $\Gamma$ the gap opens completely and the nanotube undergoes a metal-insulating transition.

\begin{figure}[htbp]
\centering
\includegraphics[width=7.5cm,height=6.0cm,angle=0.0]{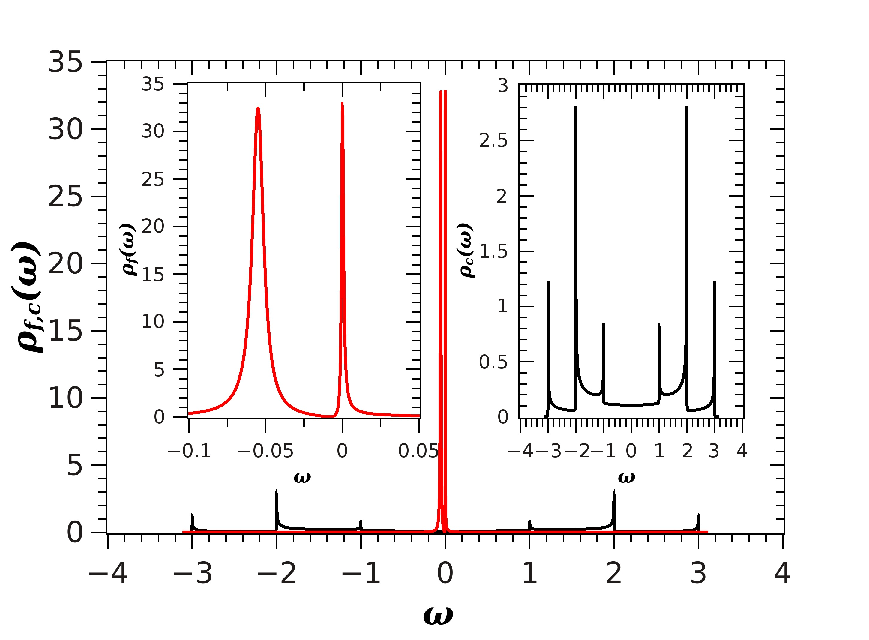}
\caption{Density of states of the localized $f$ level at $E_{f}=-5.0\Delta$ in the Kondo regime, for $\Gamma=0.0\Delta$ and temperature
$T=0.01\Delta$.} 
\label{fig3}
\end{figure}

In the Fig. (\ref{fig3}) we show the DOS of the localized level $f$ laterally coupled to a zigzag nanotube $(3,0)$, for 
$E_{f}=-5.0\Delta$ and $T=0.01\Delta$, without pression applied $(\Gamma =0)$. The figure shows the two structures that well 
characterize the Anderson impurity model, which is the widest peak left at the localized $E_{f}$ level and the narrowest 
peak formed at the chemical potential ($\mu = 0$), known as the kondo peak. On the left side of the figure we can see in 
detail the Kondo peak in chemical potential $\mu=0$  and the resonance over the localized level $E_{f}=-5.0\Delta$. On the 
right detail we show the conduction band for the SWNT.

\begin{figure}[htbp]
\centering
\includegraphics[width=7.5cm,height=6.0cm,angle=0.0]{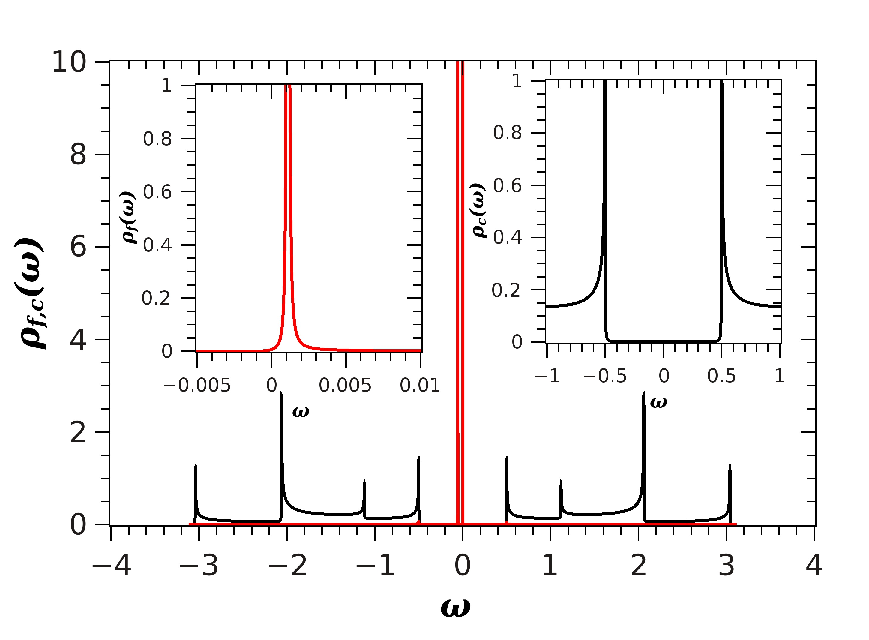}
\caption{Density of states of the localized $f$ level at $E_{f}=-5.0\Delta$ in the Kondo regime, for $\Gamma=50.0\Delta$ and temperature
$T=0.01\Delta$.}
\label{fig4}
\end{figure}

In the Fig.\ref{fig4} we use $\Gamma=50\Delta$, which is the extreme value for high pressure. The system is kept in the Kondo 
regime for $E_{f} =-5\Delta$ and $T=0.01\Delta$. With increasing $\Gamma$ values, the completeness relation Eq. (\ref{Eq.4}) 
is no longer satisfied and the Kondo peak shifts from the chemical potential. 

For large pressure values, the gap is fully formed and the Kondo effect disappears completely. The peak shown in the figure is a remnant 
of the Kondo peak. Although it still exists, it is totally out of chemical potential (see in detail of the figure) and has no contribution 
to the transport properties of the system.

\begin{figure}[htbp]
\centering
\includegraphics[width=7.5cm,height=6.0cm,angle=0.0]{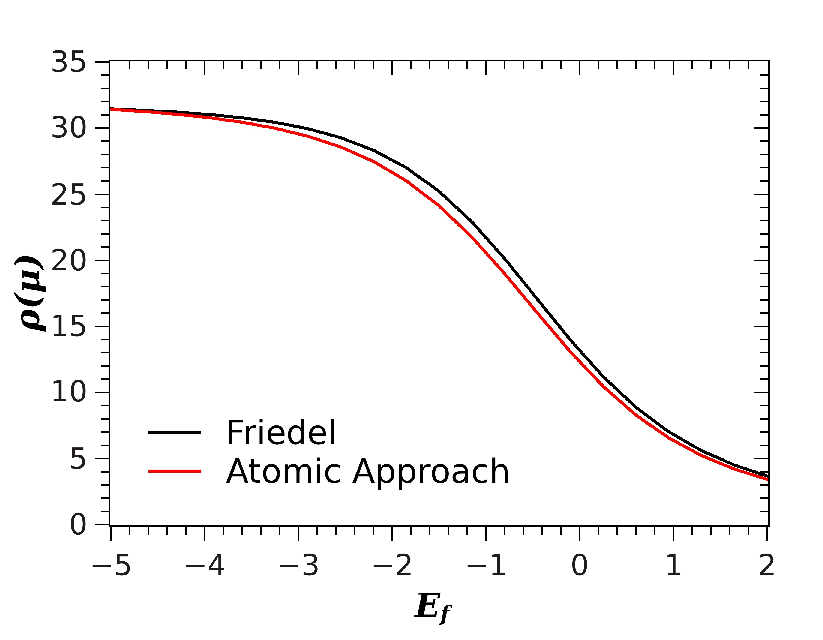}
\caption{Density os states at the chemical potential $\mu$ as a function of $E_{f}$ for the atomic approach and the Friedel 
sum rule calculated according to the Eq. (\ref{RSF})} 
\label{Densi_RSF}
\end{figure}

At zero temperature only electrons at the Fermi level, $\mu=0$, are important and the effective processes are the elastics. In this case 
the Friedel sum rule \cite{Friedel} must be satisfied

\begin{equation}
\rho_{f}(\mu)=sin^{2}\left(\frac{n_{f}\pi}{\Delta \pi}\right)
\label{RSF}
\end{equation}%
where $n_{f}$ is the occupation number of the localized electrons.
Under these same conditions and for a geometry where the impurity is laterally coupled to a ballistic channel, it is possible to show 
that the conductance can be written in a simplified way \cite{Kang2001,Torio,Aligia}:

\begin{equation}
G=\frac{2e^{2}}{h} cos^{2}(\frac{\pi}{2}n_{f})
\label{G'}
\end{equation}%
In Fig. (\ref{Densi_RSF}), for the case without applied pressure, we compare the density of states of the localized level in 
the chemical potential as a function of $E_{f}$ for the atomic approach with the values predicted by the Fridel sum rule given 
by Eq. (\ref{RSF}). We observe from the figure an excellent agreement.

\begin{figure}[htbp]
\centering
\includegraphics[width=7.5cm,height=6.0cm,angle=0.0]{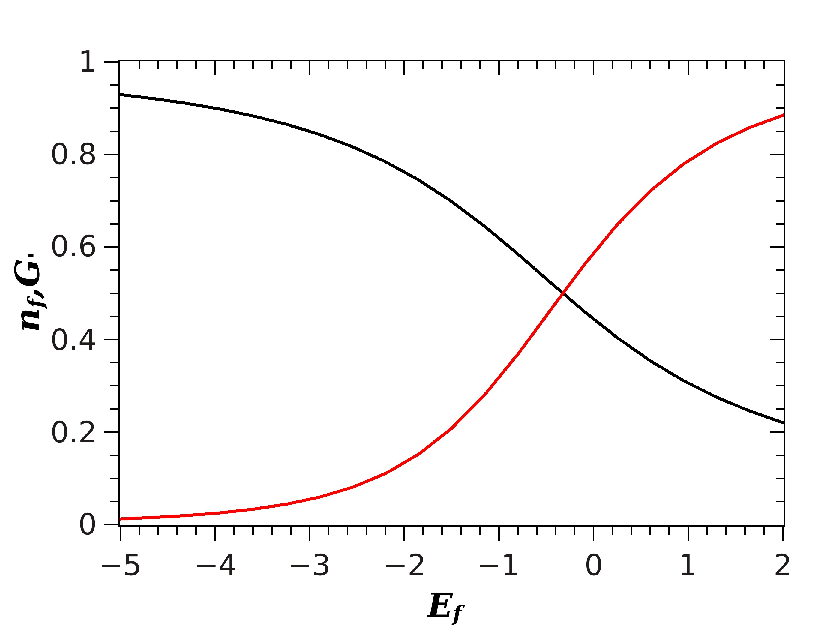}
\caption{Occupation number (black line) and conductance (red line) as a function of $E_{f}$ whith no pressure 
applied.} 
\label{fig5}
\end{figure}

In Fig. (\ref{fig5}) we present the occupation number of the localized electrons and the conductance curve as a function of $E_{f}$, 
when no pressure is applied to the system. The conductance was obtained from the exact relationship Eq. (\ref{G'}). In the Kondo regime ($E_{f}<-3.0\Delta$ and $n_{f}\rightarrow 1$), the conductance is totally suppressed by the destructive interference between the 
$c$-electrons of the nanotube and the Kondo resonance at the chemical potential.

In the empty regime ($E_{f}>0.0\Delta$ and $n_{f}\rightarrow 0$), the coupling between nanotube and impurity becomes ineffective due to 
suppression of charge and spin fluctuations. Therefore, the conductance tends to the unit value which represents the conduction of a 
ballistic channel without any interference from the Kondo effect \cite{Kang2001}.

\begin{figure}[htbp]
\centering
\includegraphics[width=7.5cm,height=6.0cm,angle=0.0]{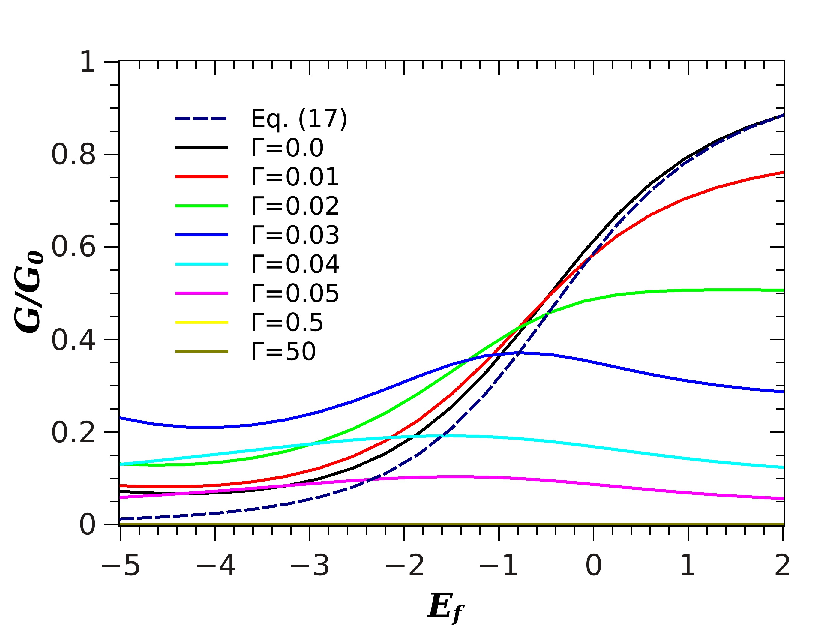}
\caption{Conductance as a function of $E_{F}$ for different values of $\Gamma$. $E_{F}$ and $\Gamma$ in units of $\Delta$ and the temperature $T=0.01\Delta$.}
\label{fig6}
\end{figure}

The Fig. (\ref{fig6}) represents the conductance as a function of the localized impurity level $E_{f}$, for different values of $\Gamma$.
For small values of pression ($\Gamma= 0.0\Delta; 0.01\Delta; 0.02\Delta$), the Kondo effect is present and the conductance is suppressed 
in the Kondo region ($E_{f}< -3.0\Delta$). As $\Gamma$ increases, the Kondo peak shifts from the chemical potential and the DOS in $\mu$ 
dramatically decreases. 

In this regime, the completeness relation is satisfied only for a smaller and smaller range of $E_{f}$ values ($<<-3.0\Delta$). For large 
values of pression ($\Gamma$=$0.5\Delta; 50\Delta$), the completeness relation can no longer be satisfied for any value of $E_{f}$. As 
the gap opens in the conduction band of the nanotube, the remnants of Kondo resonance shift from the chemical potential. There is no more 
Kondo effect acting on the system. In this case the system undergoes a transformation to an insulating regime, where the conductance is 
null for all values of $E_{f}$.

\section{Conclusions}

In this work we have employed the atomic approach \cite{Nanotech1} to study the disappearance of the Kondo effect when the hydrostatic 
pressure induces an metal-insulating transition in a zigzag SWNT. Results of the electronic density of states characterizing well the 
structure of the Kondo peak at the chemical potential, were presented. 

We have presented the suppression of the kondo effect with increasing pressure. As $\Gamma$ increases, the Kondo peak shifts from the 
chemical potential and the completeness relationship is no longer satisfied, well as calculated the occupancy number for the localized 
electrons and the result was in accordance with the Anderson impurity model regimes. We have used this result to calculate the exact 
conductance at $T = 0$ and the result was also consistent with similar systems previously studied.

We have used the Landauer formula to calculate the conductance at the coupling site between the nanotube and the magnetic impurity. 
The conductance was calculated for different pressure values, represented by the constant $\Gamma$. We have found that for very small 
values of $\Gamma$, when the pseudogap in the chemical potential is still very small, the Kondo effect interacts destructively with 
the conduction electrons of the nanotube. 

As pressure increases and the gap becomes higher, the remaining Kondo resonance peak shifts from the chemical potential and the Kondo 
effect stops acting. These studies support a more detailed investigation of the effects of the strong correlation in nanostructured 
systems in the presence of external agents, like applied pressure or electromagnetic fields.

\vspace{1.0cm}
\textbf{ACKNOWLEDGEMENTS}
We would like to express our gratitude to Professor M. S. Figueira for his encouragement and helpful discussions 
in developing this work. This work was partially supported by CNPq and CAPES (Brazilian Research Agency).

\end{document}